\documentclass[journal=jctcce,manuscript=article]{achemso}

\usepackage{bm}
\usepackage{color}
\usepackage{xspace}
\usepackage{graphicx}
\usepackage[version=3]{mhchem}
\usepackage[T1]{fontenc}
\usepackage{natbib}
\usepackage{todo}

\newcommand{\ordern}{$O(N)$\xspace}
\newcommand{\conquest}{{\sc Conquest}\xspace}

\author{Ayako Nakata}
\affiliation[NIMS]
{First-principles Simulation Group, Nano-Theory Field, International Center for Materials Nanoarchitectonics (MANA), National Institute for Materials Science (NIMS), 1-1 Namiki, Tsukuba, Ibaraki 305-0044, JAPAN}
\email{NAKATA.Ayako@nims.go.jp}

\author{Yasunori Futamura}
\email{futamura@cs.tsukuba.ac.jp}

\author{Tetsuya Sakurai}
\affiliation[TSUKUBA]
{Department of Computer Science, University of Tsukuba, 1-1-1 Tennodai, Tsukuba, Ibaraki 305-8573, Japan}
\alsoaffiliation[CREST]
{CREST, Japan Science and Technology Agency, 4-1-8 Hon-cho, Kawaguchi, Saitama 332-0012, Japan}

\author{David R Bowler}
\affiliation[UCL]
{Department of Physics \& Astronomy, University College London, Gower St, London WC1E 6BT, UK}
\alsoaffiliation[WPIMANA]
{WPI-MANA, National Institute for Materials Science (NIMS), 1-1 Namiki, Tsukuba, Ibaraki 305-0044, Japan}
\alsoaffiliation[LCN]
{London Centre for Nanotechnology, University College London, 17-19 Gordon Street, London WC1H 0AH, UK}

\author{Tsuyoshi Miyazaki}
\affiliation[NIMS]
{First-principles Simulation Group, Nano-Theory Field, International Center for Materials Nanoarchitectonics (MANA), National Institute for Materials Science (NIMS), 1-1 Namiki, Tsukuba, Ibaraki 305-0044, JAPAN}
\alsoaffiliation[UCL]
{Department of Physics \& Astronomy, University College London, Gower St, London WC1E 6BT, UK}

\title[Efficient electronic structure calculations using O(N) DFT]{Efficient calculation of electronic structure using O(N) density functional theory}

\begin{document}


\begin{abstract}
We propose an efficient way to calculate the electronic structure of large systems by combining a large-scale first-principles density functional theory code, \conquest, and an efficient interior eigenproblem solver, the Sakurai--Sugiura method. The electronic Hamiltonian and charge density of large systems are obtained by \conquest and the eigenstates of the Hamiltonians are then obtained by the Sakurai--Sugiura method. Applications to a hydrated DNA system, and adsorbed \ce{P2} molecules and Ge hut clusters on large Si substrates demonstrate the applicability of this combination on systems with 10,000+ atoms with high accuracy and efficiency.
\end{abstract}

\section{Introduction}

First-principles density functional theory (DFT) is a powerful tool to investigate the atomic and electronic structures of molecules and condensed matter.
However, most DFT calculations to date are limited in the system size which they can treat to about a thousand atoms, because the computational costs of conventional DFT calculations scale cubically to the numbers of atoms in target systems, $N$.
This size limitation makes it difficult to treat complex systems, such as biomolecules, surfaces with defects or dopants with realistic concentrations, using conventional DFT methods.

In order to overcome this size limitation, we have developed a DFT code for large-scale systems, \conquest \cite{Hernandez1996,Bowler2006,Bowler2010}, which can routinely treat systems with more than 10,000 atoms (and has been shown to scale to million atom systems \cite{Bowler2010,Arita2014}).
There are two important techniques to achieve large-scale calculations in \conquest. One is the use of ``support functions'' to express Kohn-Sham density matrices \cite{Hernandez1996}: these are real-space local orbitals having values only in finite spatial regions.
The electronic Hamiltonian and overlap matrices in this support-function basis are thus sparse, and calculations with sparse matrices are highly efficiently parallelized in \conquest \cite{Bowler2001}.  The other technique is the linear-scaling, or \ordern, method based on the density matrix minimization (DMM) method \cite{Hernandez1995,Bowler1999,McWeeny1960,Li1993}.
 In this method, the energy is minimized with respect to an auxiliary density matrix \cite{Hernandez1995}, without performing diagonalization.
 The range of the auxiliary density matrix is restricted to a finite spatial region, based on the near-sightedness principle \cite{Kohn1996}; the auxiliary density matrix is thus also sparse and the DMM calculations scale linearly with $N$.

\ordern calculations which work with the density matrix implicitly integrate over energy, and produce only the sum of the occupied eigenvalues, and not any of the eigenstates, i.e. the one-electron wavefunctions (molecular orbitals and bands) of the system.  We often want to know the individual eigenstates to analyse the electronic structure of the system, though generally within a relatively small energy range.  These can be found efficiently from the converged ground-state Hamiltonian.

There are several methods to obtain eigenstates in the interior of the spectrum of a system, such as the shift-and-invert Lanczos method \cite{Ericsson1980}, the Sakura--Sugiura method \cite{Sakurai2003}, and the FEAST algorithm \cite{Polizzi2009}. These methods enable us to avoid full diagonalization of a matrix.
The Sakurai--Sugiura method (SSM) was proposed by Sakurai and Sugiura \cite{Sakurai2003} for solving interior generalized eigenproblems of sparse matrices.
Several applications of SSM on electronic structure calculations have been reported: band structure calculations of a large system containing 10,000 silicon atoms, in conjunction with real-space finite-difference DFT \cite{Futamura2013}; molecular orbital (MO) calculations of biomolecules with the fragment molecular orbital (FMO) method \cite{Umeda2010}; and core-excited-state calculations using time-dependent DFT \cite{Tsuchimochi2008}.

The purpose of this study is to present arguments for combining \conquest and SSM. \conquest and SSM are well matched: the sparse electronic Hamiltonian can be efficiently constructed by \conquest and interior eigenproblem of this sparse matrix can be solved efficiently by SSM.
In the next section, the theoretical background of \conquest and SSM are presented. In the third section, we demonstrate that the combination enables us to investigate the electronic structure of large systems containing several thousand or several hundred thousand atoms with DFT (hydrated DNA; \ce{P2} molecules adsorbed on the Si(100) surface; and Ge hut clusters formed on the Si(001) surface). The final section gives our conclusions.

\section{Theoretical background}
\subsection{\conquest}

We briefly explain two key techniques in \conquest, the support functions \cite{Hernandez1996} and the \ordern method \cite{Hernandez1995,Bowler1999,McWeeny1960,Li1993}, to give context to the combination of \conquest with SSM.

\subsubsection{Support functions}
In DFT, the Kohn-Sham (KS) density matrix $\rho$ is expressed as
\begin{equation}
  \label{eq:1}
	\rho(\mathbf{r}, \mathbf{r}') = \sum_{n} f_n \psi_n (\mathbf{r}) \psi_n(\mathbf{r}')^*,
\end{equation}
where $\psi_n$ and $f_n$ are $n$th KS orbital and its occupation number, respectively. This would necessitate cubic scaling, so instead, in \conquest, $\rho$ is expanded using the support functions $\phi$:
\begin{equation}
  \label{eq:2}
  \rho(\mathbf{r}, \mathbf{r}') = \sum_{i \alpha,j \beta} \phi_{i \alpha} (\mathbf{r}) K_{i \alpha,j \beta} \phi_{j \beta}(\mathbf{r}')^* .
\end{equation}
Here, $\mathbf{K}$ is the density matrix in the support function basis, and $i$, $j$ and $\alpha$, $\beta$ are the indices of atoms and support functions, respectively.

Support functions are constructed as linear combinations of given basis functions $\chi$. \conquest supports two kinds of basis functions, B-spline (blip) finite element functions \cite{Hernandez1997} and pseudo atomic orbitals (PAO) \cite{Torralba2008}.
The use of blip functions enables us to improve the accuracy systematically by decreasing the width and the grid spacing of the functions, with the grid spacing being directly equivalent to a plane-wave cutoff.  The PAOs are atomic-orbital basis functions found from the pseudo-potentials and consist of radial functions ($\xi$) multiplied by spherical harmonic functions \cite{Sankey1989,Soler2002}.
Although the systematic improvement of the basis set completeness for PAOs is not straightforward, it can be improved by increasing the number of radial functions for each spherical harmonic (i.e., multiple-$\xi$ functions) as well as adding extra angular momentum shells.  The computational cost of minimisation with PAOs is at present much lower than that with blip basis functions.

Since PAOs are intrinsically localized around atoms, we can use them as support functions without modification, which we call ``primitive'' support functions.
On the other hand, we often contract the PAOs by taking linear combinations in order to reduce the number of support functions, because the computational cost increases cubically with the number of support functions \emph{per atom} in \conquest, even with the \ordern method.
Conventionally, we take a linear combination of PAOs on each atom,
\begin{equation}
  \label{eq:3}
  \phi_{i \alpha}(\mathbf{r}) = \sum_{\mu} c_{i \alpha, i \mu} \chi_{i \mu} (\mathbf{r}),
\end{equation}
where $\mathbf{c}$ are the linear-combination coefficients and $\chi_{i \mu} $ is the $\mu$th PAO on atom $i$. $\mathbf{c}$ is then optimized for all atoms in the target system.
A similar contraction method was reported by Ozaki and Kino \cite{Ozaki2003,Ozaki2004}.
Since the linear combinations are taken only over the PAOs which are centred at the target atom, the contracted ``single-site'' functions have to keep the point-group symmetry of the target atom \cite{Torralba2008}.

We have recently proposed ``multi-site'' support functions (MSSFs) \cite{Nakata2014,Nakata2015}, which are linear combinations of PAOs on the target atom and its neighbouring atoms in a finite region ($r_{\mathrm{MS}}$):
\begin{equation}
  \label{eq:4}
  \phi_{i \alpha}(\mathbf{r}) = \sum_{k}^{neighbers} \sum_{\mu \in k} C_{i \alpha, k \mu} \chi_{k \mu} (\mathbf{r}),
\end{equation}
where $k$ runs over the neighbouring atoms of atom $i$ within the region $r_{\mathrm{MS}}$. The coefficients $\mathbf{C}$ are optimized numerically \cite{Nakata2015}.
The initial values of $\mathbf{C}$ are found as localized occupied MOs using the local filter diagonalization (LFD) method \cite{Nakata2014,Rayson2009,Rayson2010}.
Because MSSFs are localised MO-like functions, we can use a number of MSSFs equivalent to a minimal basis. Therefore, by determining $\mathbf{C}$ with the LFD method and subsequently by the numerical optimization, the MSSFs can reproduce the results of the primitive functions with high accuracy while significantly reducing the computational cost.

Because the PAOs are localized functions which are restricted to finite spatial regions \cite{Sankey1989}, the support functions constructed from PAOS are also localized.
Although more delocalized than single-site support functions, MSSFs are still localized within the region ($r_{\mathrm{PAO}} + r_{\mathrm{MS}}$).
Therefore, the Hamiltonian and overlap matrices in the support-function basis are always sparse.

\subsubsection{\ordern calculations with DMM}

\conquest supports two methods to optimize the electronic states: conventional diagonalization method, scaling as $O(N^3)$ and the DMM method, scaling as \ordern.
In the diagonalization method, the density matrix $\mathbf{K}$ in (2) is obtained as the sum of the outer products of the eigenvectors of the electronic Hamiltonian $\mathbf{H}$.
On the other hand, in the DMM calculations, $\mathbf{K}$ is found by optimizing the energy with respect to an auxiliary density matrix $\mathbf{L}$.
$\mathbf{L}$ is introduced to guarantee the weak idempotency of $\mathbf{K}$ \cite{Hernandez1996,McWeeny1960,Li1993}:
\begin{equation}
  \label{eq:5}
  K = 3LSL - 2LSLSL,
\end{equation}
where $\mathbf{S}$ is the overlap matrix between support functions. The partial derivative of the total energy with respect to $\mathbf{L}$ \cite{Hernandez1996} is:
\begin{eqnarray}
  \label{eq:6}
	\frac{\partial E_{\mathrm{tot}}}{\partial L_{\alpha \beta}}
	= &[ 6 (SLH + HLS )_{\beta \alpha} \nonumber\\
          &- 4 (SLSLH + SLHLS + HLSLS)_{\beta \alpha} ].
\end{eqnarray}
The energy minimization is performed with the following restriction:
\begin{equation}
  \label{eq:7}
	L_{i \alpha, j \beta} = 0, \quad |\mathbf{R}_i - \mathbf{R}_j| > r_{\mathrm{L}},
\end{equation}
where $r_{\mathrm{L}}$ corresponds to a chosen cutoff radius. This restriction is supported by Kohn's ``near-sightedness'' principle \cite{Kohn1996}. The effect of equation~(\ref{eq:7}) is to make $\mathbf{L}$ a localized, and thus sparse, matrix.
Thus, all of the matrices in eq.~(\ref{eq:6}) are sparse so that the matrix multiplications in eq.~(\ref{eq:6}) can be performed with \ordern scaling.
The total energy obtained by the \ordern method with $r_{\mathrm{L}} = \infty$ should be equal to the one obtained by diagonalization with infinitely fine Brillouin zone sampling.

Since diagonalization is not used to solve eq.~(\ref{eq:6}), no eigenstates of the electronic Hamiltonian are calculated during the \ordern calculation.
It is usually difficult to apply conventional eigenproblem solvers to large-scale systems; however, in most practical applications, once the Hamiltonian is optimized, only a small number of eigenstates in a given energy region of interest are required.  Any eigenstate can be found in linear scaling time (as the number of basis set coefficients per eigenstate scales linearly with system size). Therefore, an efficient approach to eigenstate solution such as SSM, is an ideal choice to pair with \ordern calculations in \conquest.

\subsection{Sakurai--Sugiura method}
We now turn to a brief presentation of the Sakurai-Sugiura method.
SSM is an eigenvalue solver which computes eigenvalues located in a specified contour path $\Gamma$ on the complex plane and their associated eigenvectors.
Variants of the method using the Rayleigh--Ritz procedure \cite{SakuraiRR2007} and block versions \cite{IkegamiJCAM2010,IkegamiTJM2010} of the method have been proposed.
In this paper, we refer to these variants as the Sakurai--Sugiura method.

We denote the dimension of $\mathbf{H}$ and $\mathbf{S}$ as $n$.  The key part of SSM is a set of $M$ complex moments found as contour integrals over the contour $\Gamma$:
\begin{equation}
	W_k := \frac{1}{2 \pi \mathrm{i}} \oint_\Gamma z^k(z S - H)^{-1} SV \mathrm{d}z,
	\label{eq:ssm_ci}
\end{equation}
where $V$ is an $n \times L_\mathrm{s}$ random real matrix whose column vectors are linearly independent, $L_\mathrm{s}$ is greater than or equal to the maximum multiplicity of the eigenvalues inside $\Gamma$ and $k$ varies between $0$ and $M-1$ so that $W := [W_0, W_1, \dots, W_{M-1}]$.
By Cauchy's integral theorem, the column vectors of $W$ become linear combinations of eigenvectors corresponding to the eigenvalues located within $\Gamma$; we write $m_\Gamma$ as the number of eigenvalues (including multiplicity) inside $\Gamma$.  Then $W$ spans the same space as the eigen-subspace associated with the eigenvalues inside $\Gamma$, provided that $L_\mathrm{s} M \ge m_\Gamma$.  Either using the Rayleigh--Ritz procedure for $\mathrm{Range}(W)$, or solving a small generalized eigenvalue problem of block Hankel matrices consist of $(SV)^{\mathrm{T}} W_k$ $(k=0,1,\dots,2 M -1)$, we can compute all the eigenvalues within $\Gamma$ and their corresponding eigenvectors.

In order to approximate the contour integral eq.~(\ref{eq:ssm_ci}), we apply numerical quadrature:
\begin{equation}
	\hat{W}_k \approx W_k = \sum_{j=1}^{N_\mathrm{q}} w_j \zeta_j^k (z_j S - H)^{-1} SV,
\label{eq:numquad}
\end{equation}
where $N_\mathrm{q}$ is the number of quadrature points, $z_j$ is a quadrature point,
$w_j$ is a quadrature weight. 
The normalized quadrature point $\zeta_j$ is used instead of $z_j$ in the $k$-th order calculation of eq.~(\ref{eq:numquad}) for numerical stability.
We define $\zeta_j = (z_j - \gamma_{\mathrm{s}}) / \rho_{\mathrm{s}}$ with real scalars $\gamma_{\mathrm{s}}$ and $\rho_{\mathrm{s}} > 0$.
For instance, when $\Gamma$ is a circle, $\gamma_{\mathrm{s}}$ and $\rho_{\mathrm{s}}$ are the center and radius of the circle, respectively.
In SSM, we need to solve linear systems (with multiple right-hand sides)
\begin{equation}
	(z_j S - H) Y_j = SV \quad (j=1,2,\dots,N_\mathrm{q})
	\label{eq:linearsystem}
\end{equation}
for computing $\hat{W}_k$.

There are three levels of parallelism in the SSM algorithm.  First, eq.~(\ref{eq:linearsystem}) can be solved independently for each value of $j$.
Second, each of these linear systems can be solved in a distributed manner.  Finally, we can divide the energy range of interest into multiple regions,  and the eigenvalues in each region can be computed in parallel.
More details for the implementation and an efficient way of setting the parameters of SSM are described in Ref.~\citenum{Sakurai2013}. 

\section{Practical applications}
In order to demonstrate the usefulness of the present combination of \conquest and SSM, we results from calculations on three systems: a hydrated DNA system; \ce{P2} molecules adsorbed on the Si(100) surface; and Ge hut clusters on the Si(001) surface. Since the hydrated DNA system and \ce{P2} on Si(100), which consist of several thousand atoms, can be treated with conventional diagonalization when using MSSFs, the SCF calculations in \conquest were performed with diagonalization in these cases, allowing us to check the accuracy of the results from the SSM. For Ge clusters on Si surfaces, which consist of 10,000+ atoms, structural optimisation was performed with non-self-consistent \ordern calculations, leading to the Hamiltonian for the relaxed system. In the \conquest calculations, PBE96 generalized-gradient-approximation (GGA) exchange-correlation functional \cite{Perdew1996} and PAOs with norm-conserving pseudo potentials, which were generated with the SIESTA code \cite{Soler2002}, were used.

\subsection{Hydrated DNA}
First, we investigate the accuracy of SSM by comparing with results from exact diagonalisation, using the ScaLAPACK routine PDSYGV on the \conquest matrices.
We calculated the eigenstates of a hydrated DNA system, the B-DNA decamer 5'-d(CCATTAATGG)$_2$-3' which contains 634 atoms  in the DNA with 932 hydrating water molecules and 9 Mg counter ions, totalling 3,439 atoms. More detailed information for the geometry used is given in Ref.~\citenum{Otsuka2008}.
In \conquest, the electronic structure of the system was optimized self-consistently with the MSSFs ($r_{\mathrm{MS}}$ = 8.0 bohr). For C, N, O and P atoms, DZP (2s2p1d) PAOs \cite{noteDZP} were contracted into four support functions using the most delocalized s and p PAOs as trial vectors.
For H and Mg, DZP (2s1p) PAOs \cite{noteDZP} were contracted to one MSSF using the most delocalized s PAOs as trial vectors. The dimension of the Hamiltonian is reduced from 27,883 with DZP PAOs to 4,774 with MSSFs.
The eigenvalues and eigenvectors of the optimized Hamiltonian were obtained by both SSM and the ScaLAPACK routine to investigate both the efficiency and accuracy of SSM.
In the SSM calculation, we calculated the eigenstates in the energy range [-0.4:0.4] Hartree with an interval of 0.01 Hartree, giving 6093 states in total.
The parameters $N_\mathrm{q}$ (number of quadrature points), $M$ (number of complex moments), and  $L_\mathrm{s}$ (number of columns of the source matrix) are set to 16, 8 and 64, respectively.

\begin{figure*}
\centering
\includegraphics{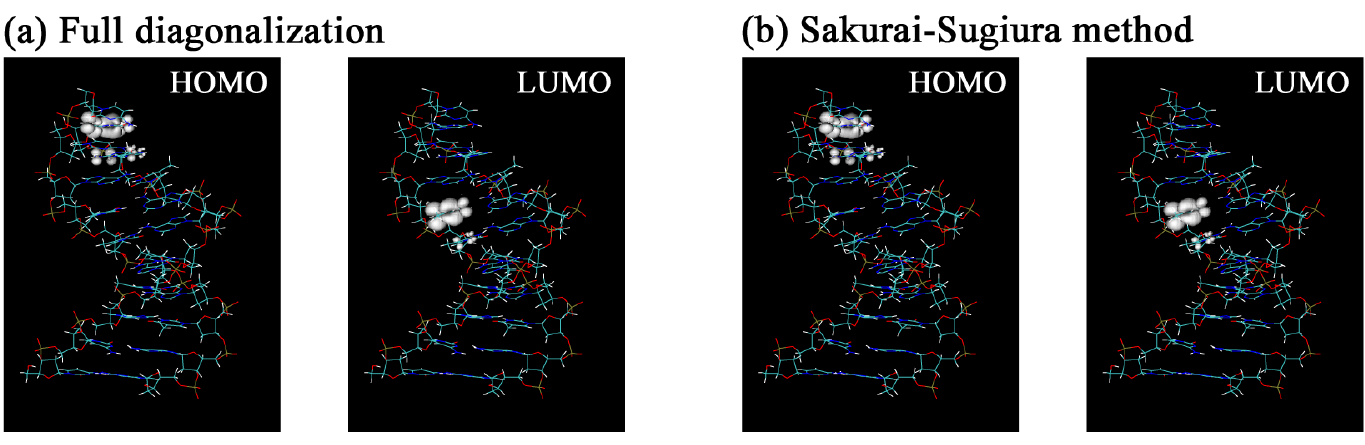}
\caption{Molecular orbital pictures of the hydrated DNA calculated by (a) full diagonalization by Scalapack and (b) Sakurai-Sugiura method.}
\label{fig:dnamo}
\end{figure*}

Table~\ref{table:dnaeig} and Figure~\ref{fig:dnamo} shows the calculated eigenvalues and eigenvectors around the Fermi level (-0.215 eV), which correspond to the energies and charge distributions of eight MOs around the HOMO-LUMO gap. Table~\ref{table:dnaeig} shows that the difference in eigenvalues calculated by SSM and those by Scalapack are negligibly small, less than $10^{-9}$ eV. The calculated HOMO-LUMO gap is 2.16 eV, which is smaller than typically-reported values for DNA systems (about 3 - 4 eV \cite{Apalkov2007}). This underestimation is likely due to the use of a GGA functional. 

\begin{table}
  \caption{Calculated eigenvalues of hydrated DNA systems [eV].}
  \begin{tabular}{lrr}
    \hline
    & \multicolumn{1}{c}{SSM} & \multicolumn{1}{c}{Scalapack} \\
    \hline
 HOMO-3 & -1.7330847154	& -1.7330847153 \\
 HOMO-2 & -1.5343553564	& -1.5343553563 \\
 HOMO-1 & -1.4458947736	& -1.4458947736 \\
 HOMO   & -1.2563299881	& -1.2563299879 \\
 LUMO   & 0.9071560149 & 0.9071560148 \\
 LUMO+1 & 0.9438977187 & 0.9438977188 \\
 LUMO+2 & 0.9673219957 & 0.9673219956 \\
 LUMO+3 & 0.9725452608 & 0.9725452608 \\
 HOMO-LUMO gap & 2.1634860030 & 2.1634860027 \\
    \hline
  \end{tabular}
  \label{table:dnaeig}
\end{table}

We can also see that the molecular orbitals (MOs) calculated by SSM and Scalapack are very similar to each other in Figure~\ref{fig:dnamo}.
The figure clearly shows that the MOs of the hydrated DNA system around the HOMO-LUMO gap are quite localized.
We suggest that the present method (i.e. the combination of \conquest and SSM) can be a powerful tool to investigate charge-transfer properties of DNA systems.

\begin{figure*}
\centering
\includegraphics[width=17cm]{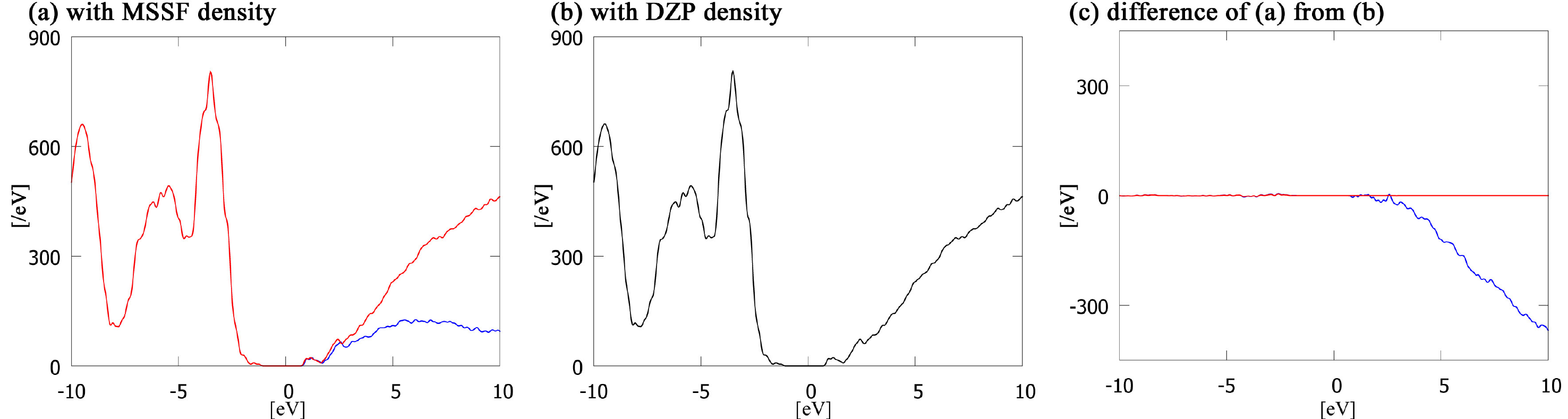}
\caption{Density of states of hydrated DNA obtained with (a) multi-site support functions (MSSF) (blue) and DZP PAOs using MSSF charge density (red) and (b) DZP PAOs using DZP SCF charge density (black). $r_{\mathrm{MS}}$ is set to be 8.0 bohr. The difference of (a) MSSF (blue) and DZP with the MSSF density (red) from (b) DZP is also shown in (c).}
\label{fig:dnados}
\end{figure*}

We also investigate the ability of the present method to improve the accuracy of the unoccupied states calculated with MSSFs.
Figure~\ref{fig:dnados} shows the density of states (DOS) of the hydrated DNA system calculated with the MSSFs (blue line in Figure~\ref{fig:dnados}a).
The results with the primitive DZP PAOs (Figure~\ref{fig:dnados}b) are also shown for comparison.
The difference between these results is shown in Figure~\ref{fig:dnados}c with the blue line, where it can clearly be seen that the difference in the occupied states is very small (the fractional difference is around 0.001) but the difference becomes larger in higher-lying unoccupied states, as already reported in Ref.~\citenum{Nakata2014,Nakata2015}.
This low accuracy in unoccupied states comes from the minimal size of the MSSF basis, and the optimisation process, which only accounts for the occupied states.  This error will be critical when we consider properties related to the unoccupied states such as excitation energies, although for most calculations it is only the accuracy of the occupied states that is  important. 

The accuracy of the unoccupied states can be improved by the following procedure: (i) optimize the charge density by SCF calculations with MSSFs; (ii) build the Hamiltonian using the primitive PAO basis, but with the optimized MSSF charge densities; and (iii) perform one-shot SS calculations to obtain the eigenvalues and eigenstates of the Hamiltonians.
Because the dimension of the reconstructed Hamiltonian in the primitive PAO basis is much larger (27,883 for the DNA system) than the Hamiltonian in MSSF basis (7,447), diagonalization in the primitive PAO basis is often difficult even if is possible in the MSSF basis.
The use of SSM (procedure (iii)) can then solve this problem.
In Figure~\ref{fig:dnados} we also show the reconstructed results: the red line in Figure~\ref{fig:dnados}a) is very similar to that of the primitive PAOs (Figure~\ref{fig:dnados}b). Figure~\ref{fig:dnados}c shows the difference from the primitive PAO results, which is almost zero throughout the eigenspectrum.

\subsection{\ce{P2} molecules on Si(100) surfaces}
In this section, we demonstrate STM simulations of large Si(100) surface samples with adsorbed \ce{P2} molecules.
Simulations of STM images and STS spectra based on DFT calculations have an important role to play in understanding the atomic and electronic structure of surfaces and adsorbates, particularly in collaboration with experiments.
In these simulations, we need only the eigenstates in limited energy regions around the Fermi levels because STM and STS measurements are only consider biases up to a few eV.
However, due to the limitation of the system size in DFT simulations, it has until now been difficult to treat large and complex surface structures.

\begin{figure}
\centering
\includegraphics[width=8cm]{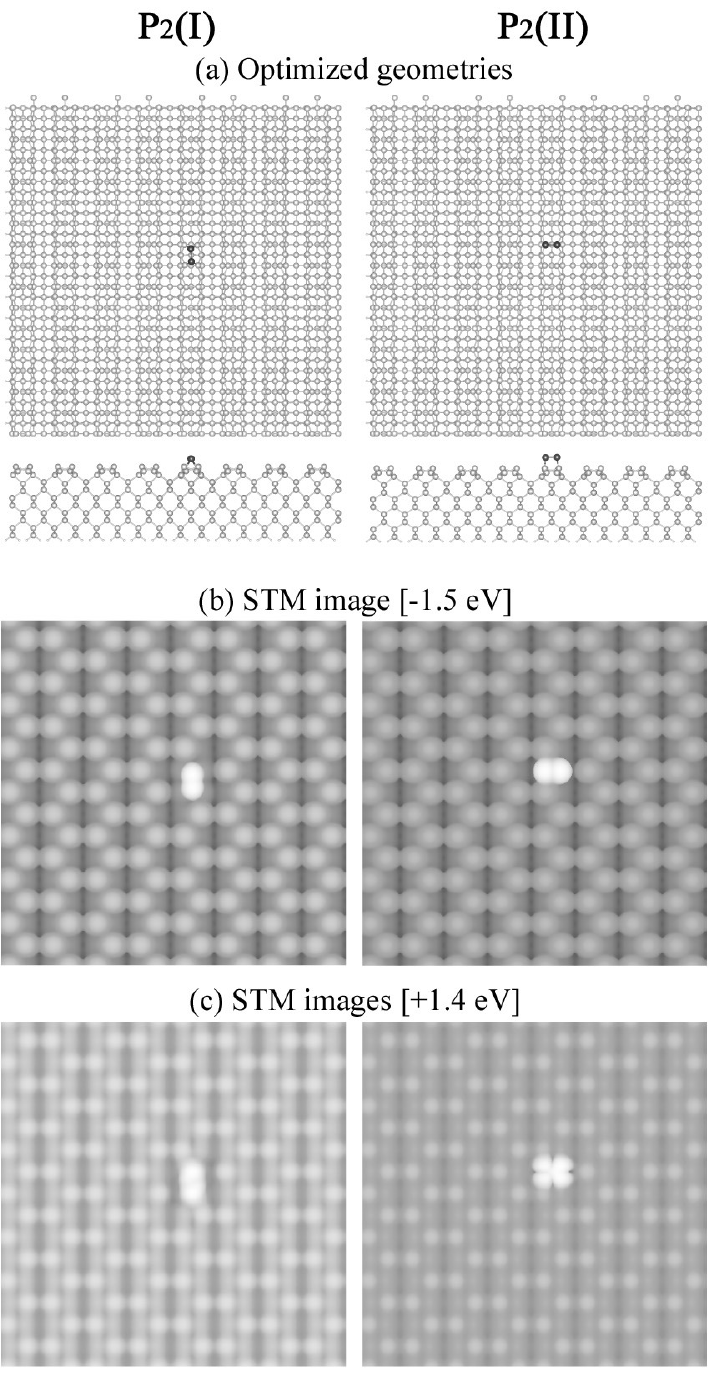}
\caption{\ce{P2} molecules on Si(100) surfaces: (a) Top and side views of the optimized geometries (Si, P and H atoms are colored in black, grey and light grey), (b) STM images at biases -1.5 eV and (c) STM images at biases +1.4 eV.}
\label{fig:surface}
\end{figure}

\ce{P2} molecules adsorbed on Si(001) tend to form two key structures, either bridging two Si dimers along the Si dimer row (\ce{P2}(I)), or just above a Si dimer (\ce{P2}(II)) \cite{Sagisaka2013}.
We modelled the \ce{P2}(I) and \ce{P2}(II) structures on large areas substrate, with ten layers of silicon, each layer consisting of 8 dimer rows each containing 16 Si dimers.
The lowest layer is terminated with H atoms. Thus, each system consists in total of 3,074 atoms (2,560 Si, 512 H and 2 P atoms).
The geometry was optimized by fixing the bottom two Si layers and H atoms. The MSSFs ($r_{\mathrm{MS}}$ = 17.0 bohr) were used for the optimization.
For Si and P atoms, four MSSFs are constructed from TZDP (3s3p2d) PAOs \cite{noteTZDP} using the most delocalized s and p PAOs as trial vectors.
The MSSFs for H atoms were constructed from DZP PAOs \cite{noteDZP} as in Section 3.1.
The optimized geometries are shown in Figure~\ref{fig:surface}(a). As reported in Ref.~\citenum{Sagisaka2013}, \ce{P2}(I) is calculated to be more stable than \ce{P2}(II) (by 0.92 eV in the present calculations). 

In order to obtain the theoretical STM images corresponding to the experimental results \cite{Sagisaka2013}, the eigenvectors only in the regions of the experimental sample biases, -1.5 eV and +1.4 eV, are required.
Therefore, we calculated the eigenstates from -1.5 eV to +1.4 eV around the Fermi energy ($e_{\mathrm{F}}$) with SSM. STM images were generated using the Tersoff-Hamann approach \cite{Tersoff1985}.
In order to improve the accuracy of the unoccupied states, as in Section 3.1, we reconstructed the Hamiltonians in primitive DZP PAO basis with the MSSF charge densities, and obtained the eigenstates by SSM subsequently.
Figures~\ref{fig:surface}(b) and \ref{fig:surface}(c) present the theoretical STM images corresponding to the sample biases -1.5 eV and +1.4 eV, which reproduce the experimental images \cite{Sagisaka2013} very well.
We would like to emphasise that only about 800 eigenstates in the interval [$e_{\mathrm{F}}-1.5$ eV: $e_{\mathrm{F}}+1.4$ eV] were needed (precisely, 776 states for \ce{P2}(I) and 778 states for \ce{P2}(II)), instead of the whole 58,924 eigenstates of the Hamiltonian which would have been required by conventional full diagonalization.
 These results suggest that our method will enable us to simulate the electronic structure of complex surfaces containing dopants and defects with realistic concentrations.

\begin{table}
  \caption{Comparison of the computation times (in second) of ScaLAPACK's PDSYGVX and SSM for the \ce{P2} on Si(001) system. $n_{\mathrm{p}}$ indicates the number of computational nodes.}
  \begin{tabular}{lrrrrrr}
  \noalign{\hrule height 1.0pt}
  $n_{\mathrm{p}}$ & 4 & 8 & 16 & 32 & 64 \\
  \hline
  PDSYGVX & 5,064 & 3,032 & 1,591 & 1,043 & 629 \\
  SSM & 924 & 469 & 239 & 134 & 85 \\
  \noalign{\hrule height 1.0pt}
  \end{tabular}
  \label{table:comp_scalapack_ssm}
\end{table}

It is also instructive to compare the computation time for SSM and ScaLAPACK's PDSYGVX subroutine for the system \ce{P2}(I).
The computational environment used for this test was the supercomputer COMA (PACS-IX) at University of Tsukuba.
Each node has Two Intel Xeon E5-2670v2 CPUs, each with 10 cores, and a clock frequency of 2.5 GHz.
While COMA has Intel Xeon Phi accelerators, we used only the CPUs for the performance evaluation.
For the parameters of SSM, we set $N_\mathrm{q}=16$, $M=8$, and $L_\mathrm{s}=100$.
We divided the interval into four equal parts.
For the numerical quadrature, we used Chebyshev points in the interval $[-1,1]$ for $\zeta_j$ and the associated barycentric weights for $w_j$ in the way presented in Ref.~\citenum{Austin2015}.
We set $\gamma_{\mathrm{s}}$ and $\rho_{\mathrm{s}}$ as the center and half width of the target energy range.
As the setting of multi-processing for SSM, we run four MPI processes per node and use five OpenMP threads per MPI process.
For PDSYGVX, we run one MPI processes per node and use 20 OpenMP threads per MPI process.
Table~\ref{table:comp_scalapack_ssm} shows the comparison of the computational times with SSM and PDSYGVX subroutine.
The table indicates that for all cases, SSM is at least five times faster than PDSYGVX, and more than seven times faster for 32 and 64 nodes.
This significant speedup comes from the high parallel efficiency of SSM, which takes advantage of the sparsity of the matrices.

\subsection{Ge hut clusters on Si(001) surfaces}
\begin{figure}
\centering
\includegraphics[width=8cm]{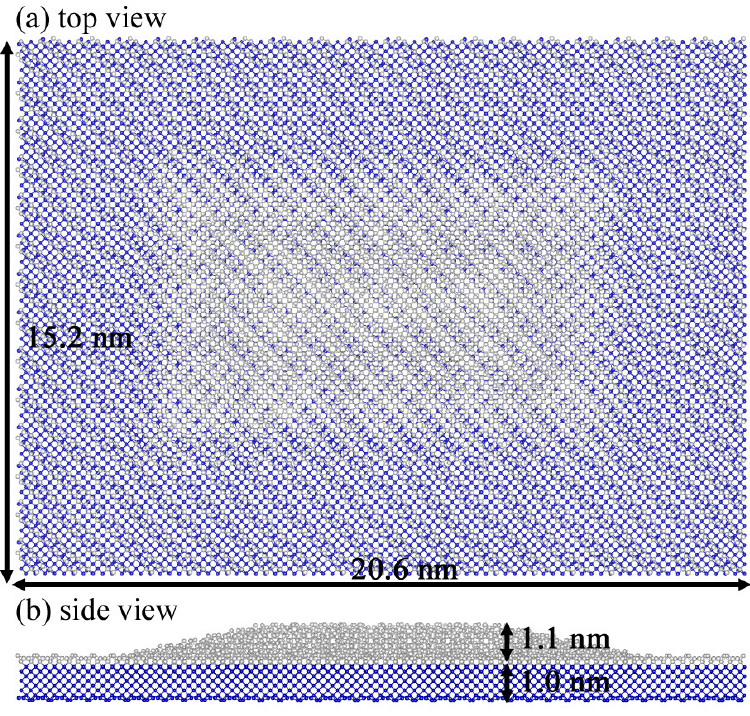}
\caption{Top and side views of the optimized geometry of the Ge hut clusters on the Si(001) (23,737 atoms).}
\label{fig:optgehut}
\end{figure}

Finally in this section, we consider systems that are out of reach of conventional DFT methods: Ge hut clusters on the Si(001) surfaces (Ge/Si(001)) consisting of (a) 23,737 and (b) 194,573 atoms. The optimized geometry of system (a) is shown in Figure~\ref{fig:optgehut}.
8$\times$13 and 20$\times$30 Ge hut-shaped clusters were placed on 14$\times$19 and 29$\times$39 SiGe substrates in systems (a) and (b), respectively. The dimensions of $\mathbf{H}$ and $\mathbf{S}$ matrices for system (a) and (b) are 213,633 and 1,751,157 with SZP PAOs.
These dimensions are beyond what could be addressed with conventional exact diagonalization.  The geometry optimization of systems (a) and (b) was perforemd with our \ordern method; the detail of the geometry optimizations will be given in a forthcoming paper \cite{ArapanArXiv}.  We used  the PZ81 LDA exchange-correlation functional \cite{Perdew1981} and SZP PAOs \cite{noteSZP}.

By combining \conquest and SSM, we can now explore not just the atomic structure but also the electronic structure of these systems.
Figure~\ref{fig:optgehut}(b) shows the sum of eigenstates in the energy window [-0.01:+0.02] eV around Fermi level (summation of 43 bands).
From the figure we can see that  the electronic charge around the Fermi level is localized at the surface of the Ge hut cluster.
Although the detailed discussion of the atomic and electronic structures of the systems, which will be investigated in a future publication \cite{ArapanArXiv}, goes beyond the purpose of this paper, the result is promising as an example of the electronic structure of one of the largest systems treated by first-principle DFT calculations.

\begin{figure}
\centering
\includegraphics[width=8cm]{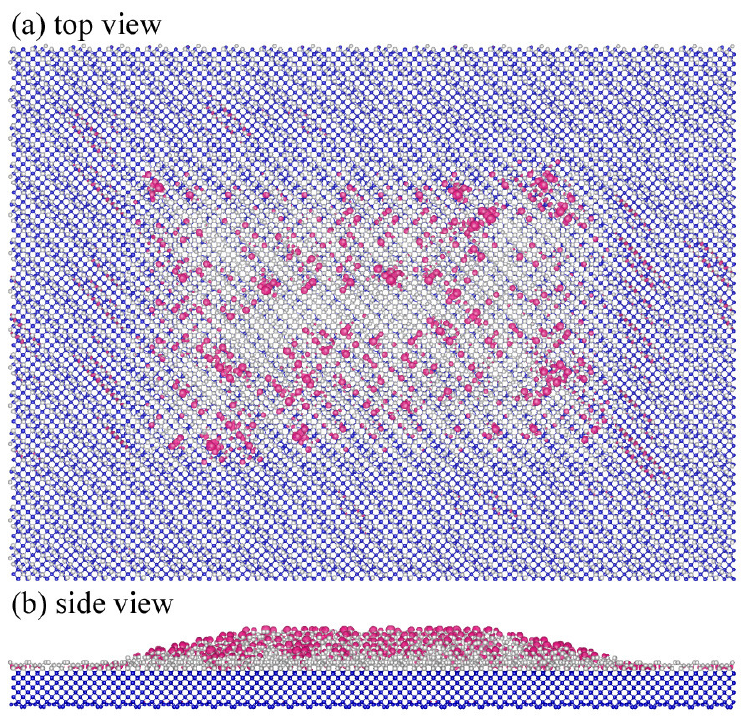}
\caption{Top and side views of the electronic density distributions of the Ge hut clusters on the Si(001) (23,737 atoms) around Fermi level.}
\end{figure}

To indicate the computation time required, we show the time for SSM calculations on both systems (a) and (b) in Table~\ref{table:sshatszp}.
For the parameters of SSM, we set $N_\mathrm{q} = 16$, $M = 8$, and $L_\mathrm{s} = 64$.
We used the K computer at RIKEN Advanced Institute for Computational Science.
The K computer has SPARC64TM VIIIfx CPUs,  with eight cores per compute node.
We used 64 nodes and 6400 nodes of the K computer for systems (a) and (b), respectively.
We run four MPI processes per node and use 8 OpenMP threads per MPI process.

\begin{table}
  \caption{The computation times of SSM for Ge hut cluster systems (a) and (b). $n_{\mathrm{p}}$ indicates the number of computational nodes.}
  \begin{tabular}{lrrr}
  \noalign{\hrule height 1.0pt}
   & \# of atoms & $n_{\mathrm{p}}$ & Time [sec] \\ 
  \hline
     System (a) & 23,737 & 64 & 146 \\ 
     System (b) & 194,573 & 6,400 & 2,399 \\
  \noalign{\hrule height 1.0pt}
  \end{tabular}
  \label{table:sshatszp}
\end{table}

\section{Conclusions}
We have shown that the combination of \conquest and the Sakurai-Sugiura method (SSM) is a powerful tool to analyze the electronic structure of very large systems.
\conquest can perform self-consistent energy minimization and geometry optimization of very large systems based on first-principles DFT by using local orbital (support) functions and a linear-scaling method. The use of the support functions makes the overlap and electronic Hamiltonian matrices sparse. The recently-developed multi-site support function method enables us to reduce the number of support functions to a minimal basis set. However, there are two key problems in analyzing the electronic structure of large systems with \conquest: the multi-site support function method gives high accuracy for occupied states but not for unoccupied states; and the linear-scaling method in \conquest does not provide the eigenvalues or eigenstates of the Hamiltonian.
SSM is a method to solve interior eigenproblems of large sparse matrices. SSM can provide the eigenvalues and eigenvectors in a finite eigenvalue range using contour integrals in complex planes, without performing diagonalization.
The combination of these methods is a powerful new tool.

We have shown that: i) the combined method is perfectly accurate in the case of a hydrated DNA system; ii) the accuracy of the unoccupied states found with multi-site support functions can be greatly improved; and iii) the method can be used to perform STM and STS simulations of complex surface structures such as \ce{P2} molecules on Si(001).  We have also shown that we can analyse the electronic structure of systems over 10,000 atoms, considering two Ge hut clusters on Si(001) surfaces with 23, 737 and 194, 573 atoms.

Our new method enables us to analyze the atomic and electronic structure of very large systems accurately and efficiently.

\begin{acknowledgement}
We gratefully acknowledge the discussion about the geometries of the DNA system, Si surfaces with \ce{P2} and the hut cluster systems with Dr. Takao Otsuka, Dr. Keisuke Sagisaka and Dr. Sergiu Arapan, respectively. This work was supported by JSPS Grant-in-Aid for Scientific Research: Grant Number 15H01052, 25286097 and 26246021, Japan. This work was partly supported by "World Premier International Research Center Initiative (WPI Initiative) on Materials Nanoarchitectonics" and "Exploratory Challenge on Post-K computer" (Frontiers of Basic Science: Challenging the Limits) by MEXT, Japan. 
Calculations were performed in the Numerical Materials Simulator at NIMS.
This research used computational resources of the K computer provided by the RIKEN Advanced Institute for Computational Science through the HPCI System Research project (Project ID:hp120170, hp140069, hp150248, hp160138).
This research in part used computational resources of COMA (PACS-IX) provided by Interdisciplinary Computational Science Program in Center for Computational Sciences, University of Tsukuba.
\end{acknowledgement}

\bibliography{Ref_JCTC_SS}

\providecommand{\latin}[1]{#1}
\makeatletter
\providecommand{\doi}
  {\begingroup\let\do\@makeother\dospecials
  \catcode`\{=1 \catcode`\}=2\doi@aux}
\providecommand{\doi@aux}[1]{\endgroup\texttt{#1}}
\makeatother
\providecommand*\mcitethebibliography{\thebibliography}
\csname @ifundefined\endcsname{endmcitethebibliography}
  {\let\endmcitethebibliography\endthebibliography}{}
\begin{mcitethebibliography}{42}
\providecommand*\natexlab[1]{#1}
\providecommand*\mciteSetBstSublistMode[1]{}
\providecommand*\mciteSetBstMaxWidthForm[2]{}
\providecommand*\mciteBstWouldAddEndPuncttrue
  {\def\EndOfBibitem{\unskip.}}
\providecommand*\mciteBstWouldAddEndPunctfalse
  {\let\EndOfBibitem\relax}
\providecommand*\mciteSetBstMidEndSepPunct[3]{}
\providecommand*\mciteSetBstSublistLabelBeginEnd[3]{}
\providecommand*\EndOfBibitem{}
\mciteSetBstSublistMode{f}
\mciteSetBstMaxWidthForm{subitem}{(\alph{mcitesubitemcount})}
\mciteSetBstSublistLabelBeginEnd
  {\mcitemaxwidthsubitemform\space}
  {\relax}
  {\relax}

\bibitem[Hern{\'{a}}ndez \latin{et~al.}(1996)Hern{\'{a}}ndez, Gillan, and
  Goringe]{Hernandez1996}
Hern{\'{a}}ndez,~E.; Gillan,~M.~J.; Goringe,~C.~M. \emph{Phys. Rev. B}
  \textbf{1996}, \emph{53}, 7147--7157\relax
\mciteBstWouldAddEndPuncttrue
\mciteSetBstMidEndSepPunct{\mcitedefaultmidpunct}
{\mcitedefaultendpunct}{\mcitedefaultseppunct}\relax
\EndOfBibitem
\bibitem[Bowler \latin{et~al.}(2006)Bowler, Choudhury, Gillan, and
  Miyazaki]{Bowler2006}
Bowler,~D.~R.; Choudhury,~R.; Gillan,~M.~J.; Miyazaki,~T. {Recent progress with
  large-scale ab initio calculations: The CONQUEST code}. Phys. Status Solidi
  Basic Res. 2006; pp 989--1000\relax
\mciteBstWouldAddEndPuncttrue
\mciteSetBstMidEndSepPunct{\mcitedefaultmidpunct}
{\mcitedefaultendpunct}{\mcitedefaultseppunct}\relax
\EndOfBibitem
\bibitem[Bowler and Miyazaki(2010)Bowler, and Miyazaki]{Bowler2010}
Bowler,~D.~R.; Miyazaki,~T. \emph{J. Phys. Condens. Matter} \textbf{2010},
  \emph{22}, 074207\relax
\mciteBstWouldAddEndPuncttrue
\mciteSetBstMidEndSepPunct{\mcitedefaultmidpunct}
{\mcitedefaultendpunct}{\mcitedefaultseppunct}\relax
\EndOfBibitem
\bibitem[Arita \latin{et~al.}(2014)Arita, Arapan, Bowler, and
  Miyazaki]{Arita2014}
Arita,~M.; Arapan,~S.; Bowler,~D.~R.; Miyazaki,~T. \emph{J. Adv. Simul. Sci.
  Eng.} \textbf{2014}, \emph{1}, 87--97\relax
\mciteBstWouldAddEndPuncttrue
\mciteSetBstMidEndSepPunct{\mcitedefaultmidpunct}
{\mcitedefaultendpunct}{\mcitedefaultseppunct}\relax
\EndOfBibitem
\bibitem[Bowler \latin{et~al.}(2001)Bowler, Miyazaki, and Gillan]{Bowler2001}
Bowler,~D.; Miyazaki,~T.; Gillan,~M. \emph{Comput. Phys. Commun.}
  \textbf{2001}, \emph{137}, 255--273\relax
\mciteBstWouldAddEndPuncttrue
\mciteSetBstMidEndSepPunct{\mcitedefaultmidpunct}
{\mcitedefaultendpunct}{\mcitedefaultseppunct}\relax
\EndOfBibitem
\bibitem[Hern{\'{a}}ndez and Gillan(1995)Hern{\'{a}}ndez, and
  Gillan]{Hernandez1995}
Hern{\'{a}}ndez,~E.; Gillan,~M.~J. \emph{Phys. Rev. B} \textbf{1995},
  \emph{51}, 10157--10160\relax
\mciteBstWouldAddEndPuncttrue
\mciteSetBstMidEndSepPunct{\mcitedefaultmidpunct}
{\mcitedefaultendpunct}{\mcitedefaultseppunct}\relax
\EndOfBibitem
\bibitem[Bowler and Gillan(1999)Bowler, and Gillan]{Bowler1999}
Bowler,~D.~R.; Gillan,~M.~J. \emph{Comput. Phys. Commun.} \textbf{1999},
  \emph{120}, 95--108\relax
\mciteBstWouldAddEndPuncttrue
\mciteSetBstMidEndSepPunct{\mcitedefaultmidpunct}
{\mcitedefaultendpunct}{\mcitedefaultseppunct}\relax
\EndOfBibitem
\bibitem[McWeeny(1960)]{McWeeny1960}
McWeeny,~R. \emph{Rev. Mod. Phys.} \textbf{1960}, \emph{32}, 335--369\relax
\mciteBstWouldAddEndPuncttrue
\mciteSetBstMidEndSepPunct{\mcitedefaultmidpunct}
{\mcitedefaultendpunct}{\mcitedefaultseppunct}\relax
\EndOfBibitem
\bibitem[Li \latin{et~al.}(1993)Li, Nunes, and Vanderbilt]{Li1993}
Li,~X.-P.; Nunes,~R.~W.; Vanderbilt,~D. \emph{Phys. Rev. B} \textbf{1993},
  \emph{47}, 10891--10894\relax
\mciteBstWouldAddEndPuncttrue
\mciteSetBstMidEndSepPunct{\mcitedefaultmidpunct}
{\mcitedefaultendpunct}{\mcitedefaultseppunct}\relax
\EndOfBibitem
\bibitem[Kohn(1996)]{Kohn1996}
Kohn,~W. \emph{Phys. Rev. Lett.} \textbf{1996}, \emph{76}, 3168--3171\relax
\mciteBstWouldAddEndPuncttrue
\mciteSetBstMidEndSepPunct{\mcitedefaultmidpunct}
{\mcitedefaultendpunct}{\mcitedefaultseppunct}\relax
\EndOfBibitem
\bibitem[Thomas~Ericsson(1980)]{Ericsson1980}
Thomas~Ericsson,~A.~R. \emph{Mathematics of Computation} \textbf{1980},
  \emph{35}, 1251--1268\relax
\mciteBstWouldAddEndPuncttrue
\mciteSetBstMidEndSepPunct{\mcitedefaultmidpunct}
{\mcitedefaultendpunct}{\mcitedefaultseppunct}\relax
\EndOfBibitem
\bibitem[Sakurai and Sugiura(2003)Sakurai, and Sugiura]{Sakurai2003}
Sakurai,~T.; Sugiura,~H. \emph{J. Comput. Appl. Math.} \textbf{2003},
  \emph{159}, 119--128\relax
\mciteBstWouldAddEndPuncttrue
\mciteSetBstMidEndSepPunct{\mcitedefaultmidpunct}
{\mcitedefaultendpunct}{\mcitedefaultseppunct}\relax
\EndOfBibitem
\bibitem[Polizzi(2009)]{Polizzi2009}
Polizzi,~E. \emph{Phys. Rev. B} \textbf{2009}, \emph{79}, 115112\relax
\mciteBstWouldAddEndPuncttrue
\mciteSetBstMidEndSepPunct{\mcitedefaultmidpunct}
{\mcitedefaultendpunct}{\mcitedefaultseppunct}\relax
\EndOfBibitem
\bibitem[Futamura \latin{et~al.}(2013)Futamura, Sakurai, Furuya, and
  Iwata]{Futamura2013}
Futamura,~Y.; Sakurai,~T.; Furuya,~S.; Iwata,~J.~I. \emph{Lect. Notes Comput.
  Sci. (including Subser. Lect. Notes Artif. Intell. Lect. Notes
  Bioinformatics)} \textbf{2013}, \emph{7851 LNCS}, 226--235\relax
\mciteBstWouldAddEndPuncttrue
\mciteSetBstMidEndSepPunct{\mcitedefaultmidpunct}
{\mcitedefaultendpunct}{\mcitedefaultseppunct}\relax
\EndOfBibitem
\bibitem[Umeda \latin{et~al.}(2010)Umeda, Inadomi, Watanabe, Yagi, Ishimoto,
  Ikegami, Tadano, Sakurai, and Nagashima]{Umeda2010}
Umeda,~H.; Inadomi,~Y.; Watanabe,~T.; Yagi,~T.; Ishimoto,~T.; Ikegami,~T.;
  Tadano,~H.; Sakurai,~T.; Nagashima,~U. \emph{J. Comput. Chem.} \textbf{2010},
  \emph{31}, 2381--2388\relax
\mciteBstWouldAddEndPuncttrue
\mciteSetBstMidEndSepPunct{\mcitedefaultmidpunct}
{\mcitedefaultendpunct}{\mcitedefaultseppunct}\relax
\EndOfBibitem
\bibitem[Tsuchimochi \latin{et~al.}(2008)Tsuchimochi, Kobayashi, Nakata,
  Imamura, and Nakai]{Tsuchimochi2008}
Tsuchimochi,~T.; Kobayashi,~M.; Nakata,~A.; Imamura,~Y.; Nakai,~H. \emph{J.
  Comput. Chem.} \textbf{2008}, \emph{29}, 2311--2316\relax
\mciteBstWouldAddEndPuncttrue
\mciteSetBstMidEndSepPunct{\mcitedefaultmidpunct}
{\mcitedefaultendpunct}{\mcitedefaultseppunct}\relax
\EndOfBibitem
\bibitem[Hern{\'{a}}ndez \latin{et~al.}(1997)Hern{\'{a}}ndez, Gillan, and
  Goringe]{Hernandez1997}
Hern{\'{a}}ndez,~E.; Gillan,~M.~J.; Goringe,~C.~M. \emph{Phys. Rev. B}
  \textbf{1997}, \emph{55}, 13485--13493\relax
\mciteBstWouldAddEndPuncttrue
\mciteSetBstMidEndSepPunct{\mcitedefaultmidpunct}
{\mcitedefaultendpunct}{\mcitedefaultseppunct}\relax
\EndOfBibitem
\bibitem[Torralba \latin{et~al.}(2008)Torralba, Todorovi{\'{c}},
  Br{\'{a}}zdov{\'{a}}, Choudhury, Miyazaki, Gillan, and Bowler]{Torralba2008}
Torralba,~A.~S.; Todorovi{\'{c}},~M.; Br{\'{a}}zdov{\'{a}},~V.; Choudhury,~R.;
  Miyazaki,~T.; Gillan,~M.~J.; Bowler,~D.~R. \emph{J. Phys. Condens. Matter}
  \textbf{2008}, \emph{20}, 294206\relax
\mciteBstWouldAddEndPuncttrue
\mciteSetBstMidEndSepPunct{\mcitedefaultmidpunct}
{\mcitedefaultendpunct}{\mcitedefaultseppunct}\relax
\EndOfBibitem
\bibitem[Sankey and Niklewski(1989)Sankey, and Niklewski]{Sankey1989}
Sankey,~O.~F.; Niklewski,~D.~J. \emph{Phys. Rev. B} \textbf{1989}, \emph{40},
  3979--3995\relax
\mciteBstWouldAddEndPuncttrue
\mciteSetBstMidEndSepPunct{\mcitedefaultmidpunct}
{\mcitedefaultendpunct}{\mcitedefaultseppunct}\relax
\EndOfBibitem
\bibitem[Soler \latin{et~al.}(2002)Soler, Artacho, Gale, Garc{\'{i}}a,
  Junquera, Ordej{\'{o}}n, and S{\'{a}}nchez-Portal]{Soler2002}
Soler,~J.~M.; Artacho,~E.; Gale,~J.~D.; Garc{\'{i}}a,~A.; Junquera,~J.;
  Ordej{\'{o}}n,~P.; S{\'{a}}nchez-Portal,~D. \emph{J. Phys. Condens. Matter}
  \textbf{2002}, \emph{14}, 2745--2779\relax
\mciteBstWouldAddEndPuncttrue
\mciteSetBstMidEndSepPunct{\mcitedefaultmidpunct}
{\mcitedefaultendpunct}{\mcitedefaultseppunct}\relax
\EndOfBibitem
\bibitem[Ozaki(2003)]{Ozaki2003}
Ozaki,~T. \emph{Phys. Rev. B} \textbf{2003}, \emph{67}, 155108\relax
\mciteBstWouldAddEndPuncttrue
\mciteSetBstMidEndSepPunct{\mcitedefaultmidpunct}
{\mcitedefaultendpunct}{\mcitedefaultseppunct}\relax
\EndOfBibitem
\bibitem[Ozaki and Kino(2004)Ozaki, and Kino]{Ozaki2004}
Ozaki,~T.; Kino,~H. \emph{Phys. Rev. B} \textbf{2004}, \emph{69}, 195113\relax
\mciteBstWouldAddEndPuncttrue
\mciteSetBstMidEndSepPunct{\mcitedefaultmidpunct}
{\mcitedefaultendpunct}{\mcitedefaultseppunct}\relax
\EndOfBibitem
\bibitem[Nakata \latin{et~al.}(2014)Nakata, Bowler, and Miyazaki]{Nakata2014}
Nakata,~A.; Bowler,~D.~R.; Miyazaki,~T. \emph{J. Chem. Theory Comput.}
  \textbf{2014}, \emph{10}, 4813--4822\relax
\mciteBstWouldAddEndPuncttrue
\mciteSetBstMidEndSepPunct{\mcitedefaultmidpunct}
{\mcitedefaultendpunct}{\mcitedefaultseppunct}\relax
\EndOfBibitem
\bibitem[Nakata \latin{et~al.}(2015)Nakata, Bowler, and Miyazaki]{Nakata2015}
Nakata,~A.; Bowler,~D.; Miyazaki,~T. \emph{Phys. Chem. Chem. Phys.}
  \textbf{2015}, \emph{17}, 31427--31433\relax
\mciteBstWouldAddEndPuncttrue
\mciteSetBstMidEndSepPunct{\mcitedefaultmidpunct}
{\mcitedefaultendpunct}{\mcitedefaultseppunct}\relax
\EndOfBibitem
\bibitem[Rayson and Briddon(2009)Rayson, and Briddon]{Rayson2009}
Rayson,~M.~J.; Briddon,~P.~R. \emph{Phys. Rev. B - Condens. Matter Mater.
  Phys.} \textbf{2009}, \emph{80}, 205104\relax
\mciteBstWouldAddEndPuncttrue
\mciteSetBstMidEndSepPunct{\mcitedefaultmidpunct}
{\mcitedefaultendpunct}{\mcitedefaultseppunct}\relax
\EndOfBibitem
\bibitem[Rayson(2010)]{Rayson2010}
Rayson,~M.~J. \emph{Comput. Phys. Commun.} \textbf{2010}, \emph{181},
  1051--1056\relax
\mciteBstWouldAddEndPuncttrue
\mciteSetBstMidEndSepPunct{\mcitedefaultmidpunct}
{\mcitedefaultendpunct}{\mcitedefaultseppunct}\relax
\EndOfBibitem
\bibitem[Sakurai and Tadano(2007)Sakurai, and Tadano]{SakuraiRR2007}
Sakurai,~T.; Tadano,~H. \emph{Hokkaido Math. J.} \textbf{2007}, \emph{36},
  745--757\relax
\mciteBstWouldAddEndPuncttrue
\mciteSetBstMidEndSepPunct{\mcitedefaultmidpunct}
{\mcitedefaultendpunct}{\mcitedefaultseppunct}\relax
\EndOfBibitem
\bibitem[Ikegami \latin{et~al.}(2010)Ikegami, Sakurai, and
  Nagashima]{IkegamiJCAM2010}
Ikegami,~T.; Sakurai,~T.; Nagashima,~U. \emph{J. Comput. Appl. Math.}
  \textbf{2010}, \emph{233}, 1927--1936\relax
\mciteBstWouldAddEndPuncttrue
\mciteSetBstMidEndSepPunct{\mcitedefaultmidpunct}
{\mcitedefaultendpunct}{\mcitedefaultseppunct}\relax
\EndOfBibitem
\bibitem[Ikegami and Sakurai(2010)Ikegami, and Sakurai]{IkegamiTJM2010}
Ikegami,~T.; Sakurai,~T. \emph{Taiwanese J. Math.} \textbf{2010}, \emph{14},
  825--837\relax
\mciteBstWouldAddEndPuncttrue
\mciteSetBstMidEndSepPunct{\mcitedefaultmidpunct}
{\mcitedefaultendpunct}{\mcitedefaultseppunct}\relax
\EndOfBibitem
\bibitem[Sakurai \latin{et~al.}(2013)Sakurai, Tadano, and
  Futamura]{Sakurai2013}
Sakurai,~T.; Tadano,~H.; Futamura,~Y. \emph{J. Algo. Comput. Tech.}
  \textbf{2013}, \emph{7}, 249--269\relax
\mciteBstWouldAddEndPuncttrue
\mciteSetBstMidEndSepPunct{\mcitedefaultmidpunct}
{\mcitedefaultendpunct}{\mcitedefaultseppunct}\relax
\EndOfBibitem
\bibitem[Perdew \latin{et~al.}(1996)Perdew, Burke, and Ernzerhof]{Perdew1996}
Perdew,~J.~P.; Burke,~K.; Ernzerhof,~M. \emph{Phys. Rev. Lett.} \textbf{1996},
  \emph{77}, 3865--3868\relax
\mciteBstWouldAddEndPuncttrue
\mciteSetBstMidEndSepPunct{\mcitedefaultmidpunct}
{\mcitedefaultendpunct}{\mcitedefaultseppunct}\relax
\EndOfBibitem
\bibitem[Otsuka \latin{et~al.}(2008)Otsuka, Miyazaki, Ohno, Bowler, and
  Gillan]{Otsuka2008}
Otsuka,~T.; Miyazaki,~T.; Ohno,~T.; Bowler,~D.~R.; Gillan,~M. \emph{J.
  Physics-Condensed Matter} \textbf{2008}, \emph{20}, 294201\relax
\mciteBstWouldAddEndPuncttrue
\mciteSetBstMidEndSepPunct{\mcitedefaultmidpunct}
{\mcitedefaultendpunct}{\mcitedefaultseppunct}\relax
\EndOfBibitem
\bibitem[not()]{noteDZP}
The range of two s, two p and a d PAOs are {(4.6, 3.4), (5.7, 3.7) and (5.7)}
  bohr for C, {(4.1, 2.9), (5.0, 3.1) and (5.0)} bohr for N, {(3.7, 2.5), (4.6,
  2.6) and (4.6)} bohr for O and {(5.2, 4.0), (6.5, 4.6) and (6.5)} bohr for P.
  Those of two s and a p PAOs are {(5.5, 4.0) and (5.5)} bohr for H and {(7.7,
  6.5) and (7.7)} bohr for Mg.\relax
\mciteBstWouldAddEndPunctfalse
\mciteSetBstMidEndSepPunct{\mcitedefaultmidpunct}
{}{\mcitedefaultseppunct}\relax
\EndOfBibitem
\bibitem[Apalkov \latin{et~al.}(2007)Apalkov, Wang, and
  Chakraborty]{Apalkov2007}
Apalkov,~V.; Wang,~X.-F.; Chakraborty,~T. \emph{{Physics Aspects of Charge
  Migration Through DNA}}; Springer Berlin Heidelberg, 2007; pp 77--119\relax
\mciteBstWouldAddEndPuncttrue
\mciteSetBstMidEndSepPunct{\mcitedefaultmidpunct}
{\mcitedefaultendpunct}{\mcitedefaultseppunct}\relax
\EndOfBibitem
\bibitem[Sagisaka \latin{et~al.}(2013)Sagisaka, Marz, Fujita, and
  Bowler]{Sagisaka2013}
Sagisaka,~K.; Marz,~M.; Fujita,~D.; Bowler,~D. \emph{Phys. Rev. B - Condens.
  Matter Mater. Phys.} \textbf{2013}, \emph{87}, 155316\relax
\mciteBstWouldAddEndPuncttrue
\mciteSetBstMidEndSepPunct{\mcitedefaultmidpunct}
{\mcitedefaultendpunct}{\mcitedefaultseppunct}\relax
\EndOfBibitem
\bibitem[not()]{noteTZDP}
The range of three s, three p and two d PAOs are {(8.5, 6.6, 4.0), (8.5, 6.6,
  4.0) and (8.5, 6.6)} bohr for Si and {(7.0, 5.6, 4.0), (7.0, 5.6, 4.0) and
  (7.0, 5.0)} bohr for P.\relax
\mciteBstWouldAddEndPunctfalse
\mciteSetBstMidEndSepPunct{\mcitedefaultmidpunct}
{}{\mcitedefaultseppunct}\relax
\EndOfBibitem
\bibitem[Tersoff and Hamann(1985)Tersoff, and Hamann]{Tersoff1985}
Tersoff,~J.; Hamann,~D.~R. \emph{Phys. Rev. B} \textbf{1985}, \emph{31},
  805--813\relax
\mciteBstWouldAddEndPuncttrue
\mciteSetBstMidEndSepPunct{\mcitedefaultmidpunct}
{\mcitedefaultendpunct}{\mcitedefaultseppunct}\relax
\EndOfBibitem
\bibitem[Austin and Trefethen(2015)Austin, and Trefethen]{Austin2015}
Austin,~A.~P.; Trefethen,~L.~N. \emph{SIAM Journal on Scientific Computing}
  \textbf{2015}, \emph{37}, A1365--A1387\relax
\mciteBstWouldAddEndPuncttrue
\mciteSetBstMidEndSepPunct{\mcitedefaultmidpunct}
{\mcitedefaultendpunct}{\mcitedefaultseppunct}\relax
\EndOfBibitem
\bibitem[Ara()]{ArapanArXiv}
Arapan S.; Bowler D. R.; Miyazaki T. A linear scaling DFT study of the growth
  of a new \{105\} facet layer on a Ge hut cluster. arXiv:1510.00526.\relax
\mciteBstWouldAddEndPunctfalse
\mciteSetBstMidEndSepPunct{\mcitedefaultmidpunct}
{}{\mcitedefaultseppunct}\relax
\EndOfBibitem
\bibitem[Perdew and Zunger(1981)Perdew, and Zunger]{Perdew1981}
Perdew,~J.~P.; Zunger,~A. \emph{Phys. Rev. B} \textbf{1981}, \emph{23},
  5048--5079\relax
\mciteBstWouldAddEndPuncttrue
\mciteSetBstMidEndSepPunct{\mcitedefaultmidpunct}
{\mcitedefaultendpunct}{\mcitedefaultseppunct}\relax
\EndOfBibitem
\bibitem[not()]{noteSZP}
The range of s, p and d PAOs are {4.4, 5.3 and 5.3} bohr for Si and {4.3, 5.4
  and 5.4} bohr for Ge.\relax
\mciteBstWouldAddEndPunctfalse
\mciteSetBstMidEndSepPunct{\mcitedefaultmidpunct}
{}{\mcitedefaultseppunct}\relax
\EndOfBibitem
\end{mcitethebibliography}



\end{document}